%% file: main.tex
\documentclass[lettersize,journal]{IEEEtran}
\usepackage{amsmath,amsfonts,amssymb,enumitem}
\usepackage{algorithmic}
\usepackage{algorithm}
\usepackage{array}
\usepackage[caption=false,font=footnotesize]{subfig}
\usepackage{textcomp}
\usepackage[nolist]{acronym}
\usepackage{stfloats}
\usepackage{comment}
\usepackage{url}
\usepackage{verbatim}
\usepackage{graphicx}
\usepackage{cite}
\usepackage{acronym}
\usepackage{xcolor}
\usepackage{units}
\input{macros}

\begin{document}
\bstctlcite{IEEEexample:BSTcontrol}

\input{acronyms}

\title{\LARGE{Performance Bounds for Velocity Estimation with Large Antenna Arrays}}

\author{\IEEEauthorblockN{Caterina~Giovannetti,~\IEEEmembership{Graduate~Student~Member,~IEEE}, \IEEEauthorblockN{Nicol\`{o}~Decarli,~\IEEEmembership{Member,~IEEE}, and Davide~Dardari,~\IEEEmembership{Senior Member,~IEEE}}
}

\thanks{C.~Giovannetti and N.~Decarli are with the National Research Council (CNR-IEIIT), and WiLab-CNIT, Bologna (BO), Italy (e-mail:  \{caterina.giovannetti, nicolo.decarli\}@ieiit.cnr.it). C.~Giovannetti is also with DEI, University of Bologna, Bologna (BO), Italy.}
\thanks{D.~Dardari is with the Dipartimento di Ingegneria dell'Energia Elettrica e dell'Informazione ``Guglielmo Marconi" (DEI), University of Bologna, and WiLab-CNIT,   Bologna (BO), Italy, e-mail: davide.dardari@unibo.it.}
}

\maketitle

\begin{abstract}
Joint communication and sensing (JCS) is envisioned as an enabler of future 6G networks. One of the key features of these networks will be the use of extremely large aperture arrays (ELAAs) and high operating frequencies, which will result in significant near-field propagation effects. This unique property can be harnessed to improve sensing capabilities. In this paper, we focus on velocity sensing, as using ELAAs allows the estimation of not just the radial component but also the transverse component. We derive analytical performance bounds for both velocity components, demonstrating how they are affected by the different system parameters and geometries. These insights offer a foundational understanding of how near-field effects play in velocity sensing differently from the far field and from position estimate.
\end{abstract}


\section{Introduction}

\IEEEPARstart{J}{oint} communication and sensing (JCS) is a technology that combines wireless data communication with radar sensing into a single system. Instead of using separate hardware and spectrum for each function, JCS enables devices to communicate and sense at the same time, often using the same signals and frequency bands. This integration improves spectrum efficiency, reduces hardware costs, opens up new applications, and it has been identified as a pillar for the forthcoming 6G \cite{ISAC:FLiu22}.
Among the quantities that can be sensed concerning a passive target, Doppler-based velocity estimation relies on the principle that the frequency of a propagating wave changes when the target and the observer are in motion relative to each other \cite{OrfB:08}.
By analyzing the frequency variations, radar systems can determine whether a target is approaching or receding and at what speed. However, only the radial velocity, i.e., the projection of the target velocity along the target-receiver direction, can produce frequency variations under traditional far-field operating conditions \cite{OrfB:08}.

Recently, communication systems are shifting towards the adoption of \acp{ELAA}  and high carrier frequencies, making traditional far-field assumptions for propagation no longer fulfilled \cite{DarDec:J21}. In the last years, this operating condition has been deeply investigated for what concerns communication and localization \cite{CuiEtAl:J23,WanEtAl:J24}. 
In the field of radar sensing, studies are still at the early stage and usually focus on the enhanced resolution offered in terms of distance/angle estimation \cite{SakEtAl:J22,SaeEtAl:J23,WanMuLiu:J23}. When dispersed antennas are employed, the ability to observe targets from different directions permits the estimation of the different velocity components using multi-static radars \cite{HeEtAl:C08}. However, thanks to the adoption of \acp{ELAA}, this capability holds even for non-dispersed antennas, allowing to estimate not only the radial component of the velocity but also the transverse one. This is the concept of \textit{velocity sensing} recently introduced in \cite{WanMuLiu:24} when employing antenna arrays working in near-field conditions, enabling to get a full picture of the target velocity and trajectory using a monostatic radar instead of multi-static radars that require challenging synchronization and expensive hardware \cite{SaeEtAl:J23}. 
Such a capability makes large antenna arrays attractive for velocity estimation, in addition to the already-known benefits of near-field operations, such as improved spatial multiplexing even in \ac{LOS} conditions \cite{DecDar:J21,CuiEtAl:J23}, the possibility of single-anchor localization \cite{GueGuiDarDju:J21} and high-resolution sensing \cite{CuiEtAl:J23,HuiEtAl:J23}. 
To the authors' knowledge, no studies are present investigating the theoretical performance limits for velocity estimation using large arrays in the near-field. The paper \cite{WanMuLiu:24} studies a maximum likelihood velocity estimator and its adoption for predictive beamforming applications, but it does not analyze the performance bounds and how the different system and geometrical parameters affect the estimation capability.

In this paper, we analytically derive the performance bounds for velocity estimation using large arrays. Specifically, we show novel results according to which \textit{(i)} the transverse velocity accuracy decreases quadratically with the target-receiver distance and \textit{(ii)} for half-wavelength spaced arrays, it does not depend on the carrier frequency. This behavior is thus completely different from that of radial velocity estimation usually considered. Moreover, we depict the impact of the different system and geometrical parameters, highlighting the differences between near-field distance estimation and transverse velocity estimation using large arrays.

\section{Signal and System Model}

\begin{figure}[t]
	\centerline{\includegraphics[width=\columnwidth]{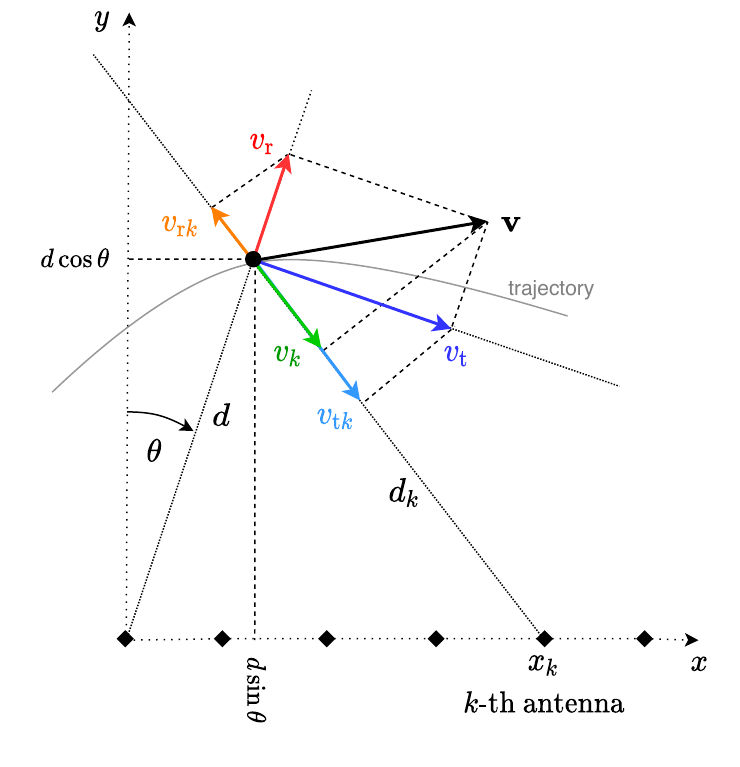}}\vspace{-0.5cm}
	\caption{Geometry of the scenario for velocity estimation with an ELAA.}
	\label{fig:NFdoppler}
\end{figure}

We consider sensing of a passive point target in a 2D scenario using a \ac{SIMO} monostatic radar at the \ac{BS}, thus with a single transmitting antenna and $\Ka$ receiving elements arranged according to a linear \ac{ELAA} deployment. 
Denote with $\delta$ the antenna spacing and let us assume $\Ka$ to be odd.  
The reference system is placed with the origin corresponding to the central element of the \ac{ELAA} at the \ac{BS}, which is oriented along the $x$-axis (see Fig.~\ref{fig:NFdoppler}).
Denote with $\tilde{\boldp}_k=[x_k, y_k]^{\top}$ the position of the $k$-th receiving antenna at the \ac{BS}, for ${k=-(\Ka-1)/2,\ldots,(\Ka-1)/2}$, so that $x_k=k \, \delta$,   $y_k=0,\, \forall k$, and $x_{0}=0$ represents the array center. 
We assume a point target placed in position ${\tilde{\boldp}=[d\sin\theta, d\cos\theta]^{\top}}$ moving with velocity $\mathbf{v}=[\vr, \vt]^{\top}$ tangent to its trajectory on the plane, 
 where $\vr$ and $\vt$ are the radial and transverse velocity components, respectively,  with respect to the center of the \ac{ELAA} (see Fig.~\ref{fig:NFdoppler}).
Let us now consider, for convenience, a polar coordinate system. 
In this case, we can write the target position as ${\boldp=[d,\theta]^\top}$, where  $d=||\tilde{\boldp}||$ denotes the distance between the target and the central element of the receiving \ac{ELAA} at the \ac{BS}, and $\theta$ denotes the angle with respect to the boresight direction of the receiving \ac{ELAA} (i.e., the \ac{AoA} of the signal received at the BS).
Define $d_k$ the distance between the target and the $k$-th receiving antenna that can be expressed as
\begin{align}   \label{eq_dk}
d_k&=\| \tilde{\boldp}-\tilde{\boldp}_k\|
= d\sqrt{1+\frac{x_k^2}{d^2}-\frac{2x_k\sin{\theta}}{d}}.
\end{align}

The central antenna element of the \ac{BS} transmits the signal $s_m(t)$ modulated at the carrier frequency $\fc$, that is
\begin{equation}\label{eq:TXsignalBP}
\tilde{s}(t)=\mathfrak{R}\left\{\sum_{m}s_m(t)  e^{\jm 2\pi\fc t}\right\}\, 
\end{equation}
where the baseband signal $s_m(t)$ is realized as an \ac{OFDM} transmission spanning $N$ subcarriers and $M$ \ac{OFDM} symbols. In particular, the $m$-th symbol component is given by
\begin{equation}\label{eq:TXsignalBB}
s_m(t)=\sum_{n}^{}\sqrt{P}\,u_{m,n}e^{\jm 2\pi n \df t} \,\rect{\frac{t-m\Ts}{\Ts}}
\end{equation}
where $P = \Ptx/N$ is the power allocated to each subcarrier, $\Ptx$ is the total transmit power, and $u_{m,n}$ is the data symbol transmitted in the $m$-th OFDM symbol and $n$-th subcarrier, with $\mathbb{E}\left\{|u_{m,n}|^2\right\}=1$, for ${m=-(M-1)/2,\ldots,(M-1)/2}$ and ${n=-(N-1)/2,\ldots,(N-1)/2}$. We denote with $\mathbb{E}\{ \cdot \}$ the statistical expectation.
The symbol time is $\Ts=T+\Tcp$ where $\Tcp$ stands for the cyclic prefix duration and $T=1/\df$, with $\df$ indicating the \ac{SCS}.

After classical cyclic prefix removal and FFT processing, the received signal is
\begin{align}\label{eq:RXsigNF}
r_{m,n,k} &=y_{m,n,k}+ z_{m,n,k}  \\
&=\sqrt{P} \beta_{m,n,k} u_{m,n}  e^{-\jm 2 \pi f_n \tau_k} e^{\mathrm{j}2 \pi \nu_{n,k} m\Ts}  + z_{m,n,k} \nonumber
\end{align}
where $f_n \triangleq \fc + n\df$ is the frequency associated with the $n$-th subcarrier, $\beta_{m,n,k}$ is the channel gain coefficient,  $\tau_{k}=\frac{d+d_k}{c}$ is the round-trip delay associated to the $k$-th receiving antenna, and ${\nu_{n,k}=\frac{f_n}{c} \left(\vr+\mathbf{v}\cdot\ek\right)}$ is the round-trip Doppler shift, with $\mathbf{a} \cdot \mathbf{b}$ indicating the scalar product between vectors $\mathbf{a}$ and $\mathbf{b}$, and $c$ the speed of light. Here, the term $\ek={\mathbf{d}_k}/{|| \mathbf{d}_k||}$ is a unit norm vector denoting the direction between the $k$-th receiving antenna element and the target, so that $\nu_{n,k}$ includes a component which is the projection of the target's velocity along this direction for each receiving antenna. 
The term $z_{m,n,k}$ indicates the \ac{AWGN}, with $z_{m,n,k}\sim\mathcal{CN}(0,\sigma^2)$ and ${\sigma^2=\kb T_0 F\df}$, being $\kb$ the Boltzmann constant, $T_0$ the reference temperature, and $F$ the receiver noise figure.
We consider the channel gain equal for all the antennas, which is reasonable for the practical size of the receiving \ac{ELAA} and small relative bandwidth $B=N\df\ll\fc$, so that we can write ${\beta_{m,n,k}=\beta,\, \forall m,n,k}$. 

Starting from the received signal \eqref{eq:RXsigNF}, the \ac{BS} aims at estimating the position $\boldp$ and velocity $\mathbf{v}$ through the estimation of their components   $[d , \theta]$ and $[\vr, \vt]$, respectively, thanks to the use of the ELAA.

\section{Velocity Estimation Performance Bounds}\label{sec:Bound}

The optimal way of processing the receiving signal would require the joint estimation of all the unknown parameters $\mathbf{\Theta}=\{d,\theta,\vr,\vt\}$ from the ${M \times N \times \Ka}$ observations. This would imply a very high complexity, especially when operating in the near-field region where the planar wavefront approximation does not hold.
To focus here on the estimation of the radial and transverse velocity components when adopting the \ac{ELAA}, we assume that the distance $d$ and the \ac{AoA} $\theta$ have already been estimated, as done also in \cite{WanMuLiu:24}. This can be realized by exploiting the information coming from the different subcarriers (e.g., for distance estimation), and/or from the phase profile along the array caused by the spherical wavefront (e.g., for distance and angle estimation) \cite{WanMuLiu:J23}. 
Thus, starting from this assumption, we want to evaluate the performance limits on the estimation \ac{MSE} of the radial and transverse velocity components $\vr$ and $\vt$. Since the transmitted symbols are known by the receiver and used as pilots, in the following we consider $u_{m,n}=1$ in \eqref{eq:RXsigNF}; moreover, $\tau_k$ is known since the distance $d$ and the angle $\theta$ have been assumed already estimated.

The velocity estimation quality in terms of \ac{MSE} is lower bounded by the \ac{CRLB} \cite{Kay:B93}.
In this case, we have two scalar parameters to consider for the \ac{CRLB} analysis, that are ${\mathbf{\Theta} = \{\vr, \vt\}}$. The $(i,j)$-th element of the \ac{FIM} can be obtained as \cite{Kay:B93}
\begin{equation}\label{eq:FIM}
[\mathbf{J}]_{i,j}= \frac{2}{\sigma^2} \Re\left\{\sum_{m,n,k}\left[\frac{\partial y_{m,n,k}}{\partial\mathbf{\Theta}_i} \right]^* \left[\frac{\partial y_{m,n,k}}{\partial\mathbf{\Theta}_j} \right]  \right\}
\end{equation}
where $y_{m,n,k}$ is the noise-free version of $r_{m,n,k}$ in \eqref{eq:RXsigNF}. Thus we have a $2 \times 2$ FIM in the form
\begin{equation}
\mathbf{J}=  
\begin{bmatrix}
\Jrr & \Jrt \\
\Jtr & \Jtt  \\
\end{bmatrix}
\end{equation}
with $\Jrt=\Jtr$, and the corresponding \acp{CRLB} on the radial and transverse velocities are given by, respectively,
\begin{align}
\mathsf{CRLB}^{(\vr)}=\frac{1}{\operatorname{det}\mathbf{J}}\Jtt \, , \quad \quad
\mathsf{CRLB}^{(\vt)}=\frac{1}{\operatorname{det}\mathbf{J}}\Jrr \label{eq:CRLB}
\end{align}
where ${\operatorname{det}\mathbf{J}}=\Jrr\Jtt-\Jtr^2$.

According to the geometry of Fig.~\ref{fig:NFdoppler}, we can project the radial and transverse velocity components along the direction $\ek$ by obtaining the projections $\vrk$ and $\vtk$, respectively. It holds
\begin{equation}\label{eq:vk}
    \vk=\vrk+\vtk
\end{equation}
where $\vk$ is the projection of the velocity $\mathbf{v}$ along the direction $\ek$.
Then, we can write
\begin{equation}\label{eq:nuk}
    \nu_{n,k}=\frac{f_n}{c}(\vr+\vk) .       
\end{equation}
The projections $\vrk$ and $\vtk$ can be obtained as \cite{WanMuLiu:24}
\begin{align}   
    \vrk=\,\,&q_k \vr \label{eq:vdk}\\
    \vtk=\,\,&p_k \vt \label{eq:vtk}
\end{align}
where 
\begin{align}   
    q_k =& \frac{d-x_k\sin\theta}{d_k} = \frac{1-\frac{x_k\sin\theta}{d}}{\sqrt{1+\frac{x_k^2}{d^2}-\frac{2x_k\sin{\theta}}{d}}} \label{eq:qk}\\
    p_k =& \frac{x_k\cos\theta}{d_k} = \frac{x_k\cos\theta}{d\sqrt{1+\frac{x_k^2}{d^2}-\frac{2x_k\sin{\theta}}{d}}} \label{eq:pk} \, .
\end{align}

By substituting \eqref{eq:vdk} and \eqref{eq:vtk} in \eqref{eq:vk}, and considering \eqref{eq:nuk} in the model \eqref{eq:RXsigNF}, we can compute the derivatives in \eqref{eq:FIM} and obtain the components of the \ac{FIM}
\begin{align}
    \Jrr &= \sum_n I_n \sum_k \left ( 1 + q_k\right )^2 \label{FIM_vd}
\end{align}
\begin{align}
    \Jtt &= \sum_n I_n \sum_k p_k^2  \label{FIM_vt}
\end{align}
\begin{align}
\Jtr &= \sum_n I_n \sum_k p_k \left ( 1 + q_k\right ) 
\end{align}
where 
\begin{equation}
    I_n=\frac{2 \pi^2 \, f_n^2\, M \, \SNR  \,  (M^2 -1)\Ts^2 }{3c^2}
\end{equation}
and we defined the \ac{SNR} as ${\SNR=P\beta^2/\sigma^2}$.
As it will be evident in the numerical results, for practical configurations in terms of antenna size, it holds ${\Jrr \Jtt \gg \Jtr^2}$ so that ${{\operatorname{det}\mathbf{J}}\approx \Jrr\Jtt}$ and we have ${\mathsf{CRLB}^{(\vr)}\approx{1}/{\Jrr}}$ and ${\mathsf{CRLB}^{(\vt)}\approx{1}/{\Jtt}}$.

The limit for $d\rightarrow\infty$ in \eqref{eq:qk} is $\qk\rightarrow1$ so that the right-hand summation in \eqref{FIM_vd} tends to $4\Ka$. Thus, for the estimation of the radial velocity component $\vr$, the information \eqref{FIM_vd} tends to the inverse of the traditional \ac{CRLB} for velocity estimation in far-field conditions, which is given by \cite{OFDM_Radar_Phd_2014}
\begin{align}
    \mathsf{CRLB}^{(\vr)} &= \frac{3c^2} {8 \pi^2 \, \fc^2\, M N \Ka \SNR  \,  (M^2 -1)\Ts^2 }  \label{CRB_vr} \, 
\end{align}
where we considered $f_n\approx\fc$ since $B\ll\fc$. Notice that $(M^2 -1)\Ts^2$ is approximately the squared signal duration $\Tobs^2$, and in \eqref{CRB_vr} it  is highlighted the SNR gain $M N \Ka$ obtained thanks to the exploitation of $M$ symbols, $N$ subcarriers and $K$ antennas for the estimation.
When ${\theta=0}$, \eqref{FIM_vd} becomes
\begin{align}
    \Jrr \!\!=\!\! \frac{2 \pi^2 \! \fc^2 M N \SNR    (M^2 \!-\!1)\Ts^2 }{3c^2}  \!\sum_k\! \left ( \!1 \!+\! \frac{1}{\sqrt{1 \!+\! k^2\frac{\delta^2}{d^2} }}\!\right )^{\!\!2}\!\!. \label{FIM_vd2} 
\end{align}
Interestingly, the information on the target's radial velocity reduces as the distance $d$ from the array decreases, since $k^2\delta^2/d^2>0$, resulting in an SNR gain lower than $K$ for a small distance. However, for practical array dimensions and operating distances (in particular, when the distance becomes significantly higher than the largest $x_k$, i.e., than the array aperture ${\Da=(K-1)\delta}$), the term $k^2\delta^2/d^2$ is small. Therefore, as far as the distance $d$ exceeds the array aperture $\Da$, we have $1/\Jrr\rightarrow \mathsf{CRLB}^{(\vr)}$ in \eqref{CRB_vr} which does not depend on the distance $d$ and on the array aperture $\Da$.

For what concerns the transverse velocity,  its estimation at a large distance is not possible; in fact, we have $\pk\rightarrow0$ for $d\rightarrow\infty$ (in practice, when $d$ is much larger than the array aperture $\Da$) so that $\Jtt\rightarrow0$ and no information can be retrieved. Differently, when the array aperture increases and/or the distance decreases, we have $\pk\neq 0$, and transverse velocity estimation becomes feasible. The information on transverse velocity is maximum on the boresight direction of the \ac{ELAA} according to \eqref{FIM_vt} and \eqref{eq:pk} (i.e., $\theta=0$); differently, it is not possible to estimate the transverse velocity when $\theta=\pm\frac{\pi}{2}$.  In fact, in this case, the direction $\ek$ is the same for all the antennas and corresponds to the radial direction, so no further information becomes available rather than the radial velocity. The same behavior is experienced in near-field distance estimation with \acp{ULA}, which is not possible for $\theta=\pm\frac{\pi}{2}$ \cite{KorEtAl:J10}.
For the case of transverse velocity estimation, we can write for $\theta=0$
\begin{align}
    \Jtt \!=\! \frac{2\pi^2\,  \fc^2 \,  M N \SNR  \,  (M^2 \!-\! 1)\Ts^2 \delta^2}{3c^2\,d^2} \sum_k \frac{k^2}{1 \!+\! k^2\frac{\delta^2}{d^2} }  \label{FIM_vt2}
\end{align}
which decreases with the square of the distance $d$. Considering the practical case where the target is not very close to the \ac{BS}, the term $k^2\delta^2/d^2$ is generally small, and the following approximate expression is obtained
\begin{align}
    \Jtt &\approx \frac{\pi^2\,  \fc^2 \,  M N \Ka \SNR  \,   (M^2 - 1)\Ts^2  (\Ka^2-1)\delta^2}{18c^2\,d^2} \nonumber \\
 &\approx \frac{\pi^2 \, \fc^2\, M N \Ka \SNR  \,  \Tobs^2 \Da^2 }{18 c^2 d^2}  \, .
    \label{FIM_vt3}
\end{align}

As for the traditional \ac{CRLB} on radar-based velocity estimation in \eqref{CRB_vr}, the information on the transverse velocity is proportional to the squared signal duration. Larger carrier frequency $\fc$ is beneficial for both radial and transverse velocity estimation. The adoption of a proper array aperture $\Da$ can be traded to achieve the required transverse velocity accuracy depending on the operating distance $d$, while the \ac{SNR} and the symbol time have the same impact on both the components.

\begin{figure}[t]
\vspace{-0.2cm}
	\centerline{\includegraphics[width=\columnwidth]{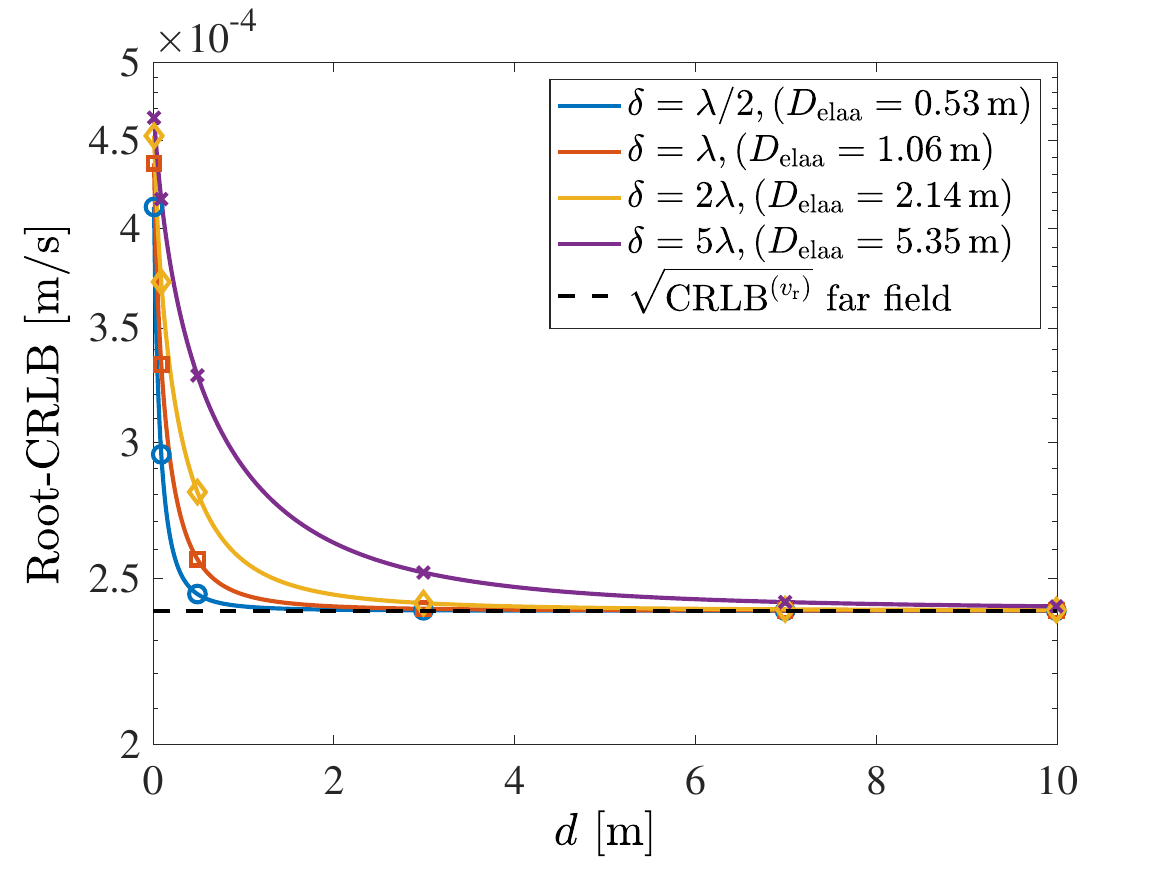}}
	\caption{Root-CRLB for the radial velocity as a function of the distance $d$ for different apertures $\Da$. Markers stand for $\sqrt{1/\Jrr}$ in the same setting.}
	\label{fig:Fig1}
\end{figure}

Expression \eqref{FIM_vt3} shows a deep difference between what happens in the near field for distance estimation and transverse velocity estimation. In fact, in near-field distance estimation, the CRLB is proportional to $d^2/\fc^2\Da^4$ \cite{KorEtAl:J10}; as a result, in that case, the root-\ac{CRLB} is inversely proportional to the Fraunhofer distance ${d_{\mathrm{ff}}=2\Da^2/\lambda}$, with $\lambda=c/\fc$. Therefore, increasing the array aperture $\Da$ (i.e., number of antennas, antenna spacing) and/or the carrier frequency is beneficial for the estimation quality, with a larger impact of the array aperture. Instead, in transversal velocity estimation using \acp{ELAA}, the parameters affecting the quality are the same according to \eqref{FIM_vt3}, but they influence the bound all with the second power. Differences are evident when considering a half-wavelength spaced array ($\delta=\lambda/2$). In this case, the information on the transverse velocity is
\begin{align}
    &J_{\vt \vt} \approx \frac{\pi^2 \,  M N \Ka \SNR  \,   (M^2 - 1)\Ts^2 (\Ka^2-1)}{72d^2} \, \label{FIM_vt4}
\end{align}
which 
does not depend on the carrier frequency, differently from the radial velocity estimation or what happens for near-field distance estimation. These results suggest that the possibility of gathering information on the transverse velocity is not strictly related to the near-field behavior intended as spherical wavefront propagation, rather than on the geometrical aspects enabling to project the target velocity along the set of directions corresponding to the different antennas of the array. 

\section{Numerical Results}

We now present some results concerning the quality of velocity estimation using a linear \ac{ELAA}. If not differently specified, we consider $\fc=\unit[28]{GHz}$, $\Ka=\unit[101]{}$ antennas, $N=\unit[1]{}$ and $\SNR=\unit[0]{dB}$. A packet with $M=14$ OFDM symbols is considered, and symbol time $\Ts=\unit[16.6]{ms}$. According to the expressions reported in Sec.~\ref{sec:Bound}, the presented root-CRLB curves will scale with $\sqrt{N}$ when considering $N>1$. 

Fig.~\ref{fig:Fig1} shows the root-\ac{CRLB} for the radial velocity estimation in \eqref{eq:CRLB}-left as a function of the distance $d$, for a small distance from the array. Markers indicate the approximation obtained as $\sqrt{1/\Jrr}$, with $\Jrr$ given by \eqref{FIM_vd2}, showing that it is very tight for the condition of interest. The traditional far-field \ac{CRLB} according to \eqref{CRB_vr} is reported for comparison.  Results are given for different apertures $\Da$, but with the same number of antennas $\Ka$ in order to consider a constant overall received power. As it is possible to notice, a small deviation from the far-field \ac{CRLB} is experienced only at very small distance from the array, in particular, for a distance below the array aperture $\Da$, region also known as \textit{geometric near field} \cite{DecDar:J21}.

\begin{figure}[t]
\vspace{-0.2cm}
	\centerline{\includegraphics[width=0.99\columnwidth]{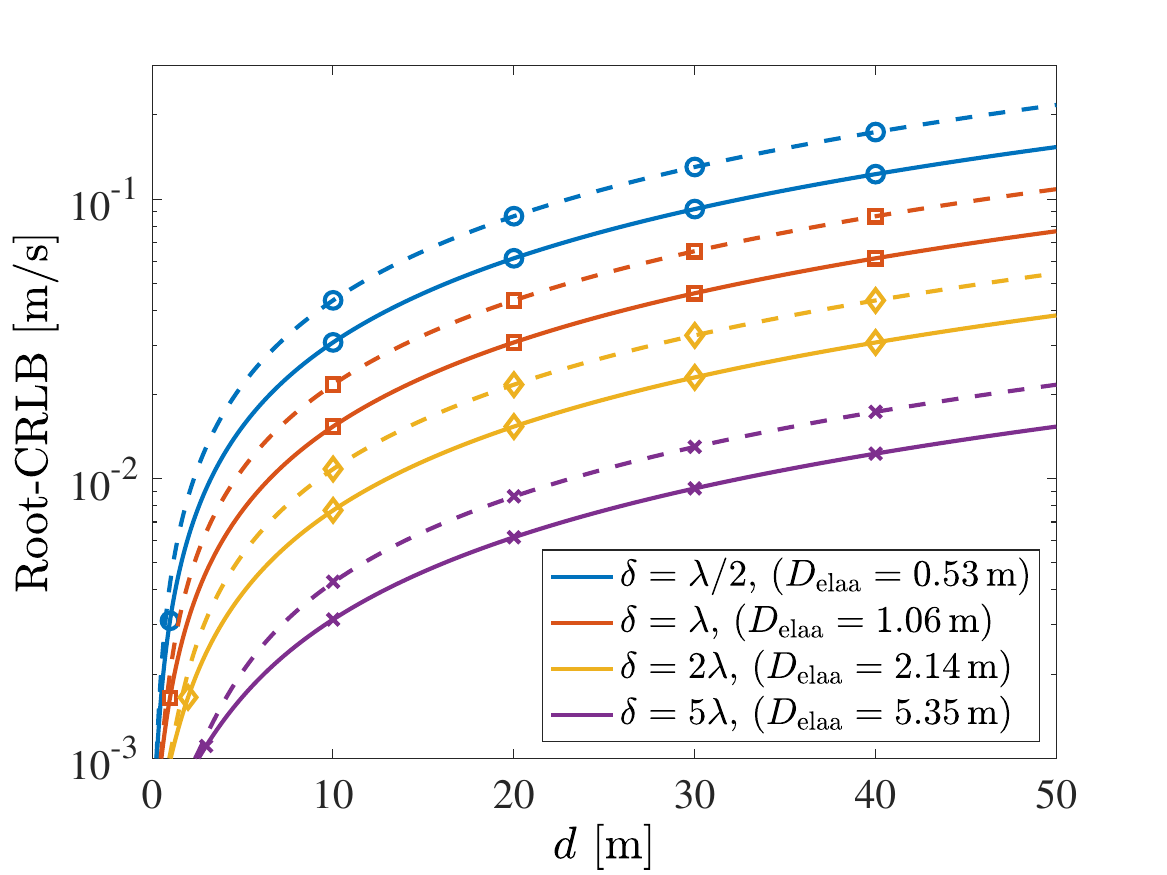}}
	\caption{Root-CRLB for the transverse velocity as a function of the distance $d$ for different apertures $\Da$. Continuous lines (--) are for $\theta=0$; dashed lines (-$\,$-) are for  $\theta=45^{\circ}$. Markers stand for $\sqrt{1/\Jtt}$ in the same setting.}
	\label{fig:Fig2}
\end{figure}

In Fig.~\ref{fig:Fig2} the root-\ac{CRLB} for the transverse velocity in \eqref{eq:CRLB}-right is reported as a function of the distance $d$, considering $\theta=0$ and $\theta=45^\circ$. Again, the comparison between the exact root-\ac{CRLB} and its approximation  $\sqrt{1/\Jtt}$ (markers) shows a very good agreement. It can be noticed that the accuracy decreases as the distance $d$ increases, as well as when the angle increases. The array aperture $\Da$ has a fundamental role in achieving a good estimation quality.

\begin{figure}[t]
	\centerline{\includegraphics[width=\columnwidth]{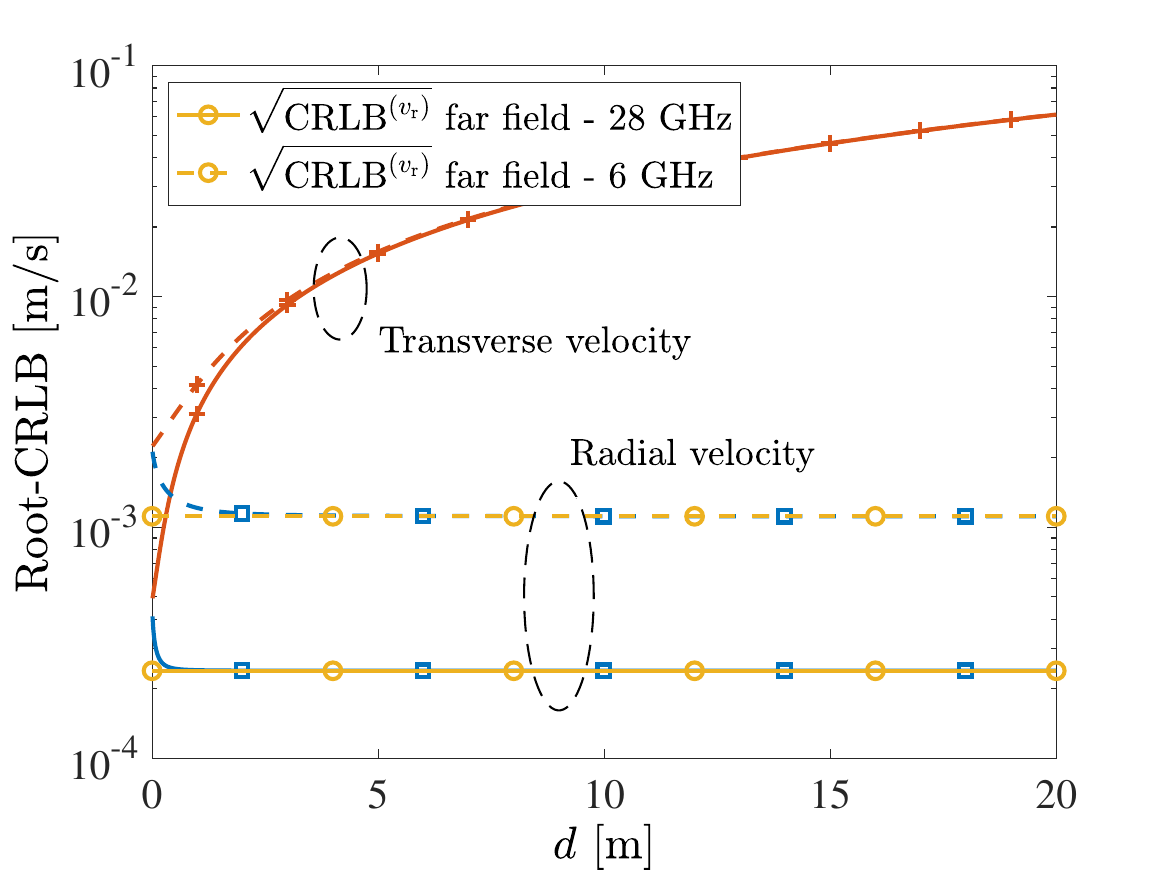}}
	\caption{Root-CRLB for radial ($\square$) and transverse ($+$) velocities as a function of the distance $d$ for different carrier frequencies and $\delta=\lambda/2$ (angle ${\theta=0}$). Continuous lines (--) are for $\fc=\unit[28]{GHz}$; dashed lines (-$\,$-) are for  ${\fc=\unit[6]{GHz}}$. Comparison with the far-field CRLB in \eqref{CRB_vr}.} 
	\label{fig:Fig3}
\end{figure}

Fig.~\ref{fig:Fig3} compares accuracy related to the radial velocity (blue curves with $\square$) and to the transverse velocity (red curves with $+$)  in terms of root-CRLB as a function of the distance $d$ for a fixed number of antennas $K$, a half-wavelength antenna spacing $\delta=\lambda/2$ and two different carrier frequencies, i.e., $\fc=\unit[28]{GHz}$ (continuous lines) and $\fc=\unit[6]{GHz}$ (dashed lines). Thus, the two conditions translate in two different array apertures $\Da$.
It is possible to notice that the radial accuracy is close to the traditional CRLB (yellow curves with $\circ$), and it is always larger than the transverse one. In fact, by comparing \eqref{CRB_vr} and \eqref{FIM_vt3} it can be seen that when $d=\Da/(4\sqrt{3})$, the estimation quality for the radial velocity equals the estimation quality for the transverse one, at $\theta=0$. 
Thus, an operating distance smaller than the array aperture $\Da$ is required to achieve a balance in terms of estimation quality among the different components of the velocity. 
Moreover, the radial velocity accuracy slightly deteriorates when close to the array. The transversal velocity accuracy does not depend on the carrier frequency, since we are comparing the results considering half-wavelength antenna spacing; differently, the estimation quality for the radial velocity improves as the carrier frequency increases.

\begin{figure}[t]
	\centerline{\includegraphics[width=\columnwidth]{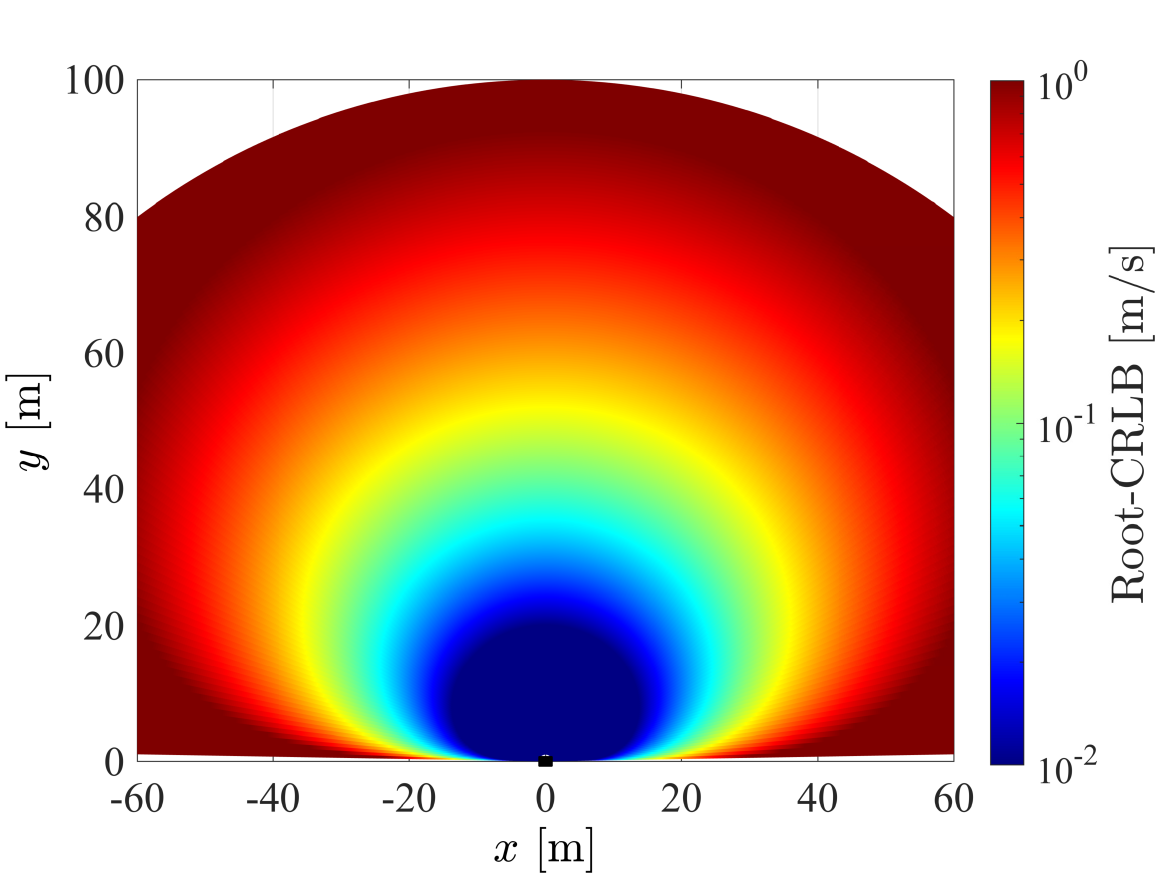}}
	\caption{Distribution of the root-CRLB for transverse velocity on the 2D plane.}
	\label{fig:Fig4}
\end{figure}

Finally, a 2D map considering the distribution of the root-CRLB for the transverse velocity in the semi-plane in front of the \ac{ELAA} is reported in Fig.~\ref{fig:Fig4}. In this case, we considered a transmit power $\Ptx=\unit[23]{dBm}$ and a receiver noise figure $F=\unit[9]{dB}$,  to include the impact of the path loss in the \ac{SNR}. A target with radar cross section of $\unit[1]{m^2}$ is assumed, and $\unit[0]{dBi}$ gain antennas. As evident, the transversal accuracy decreases as the distance from the array increases due to the combined effect of the term $d^2$ in \eqref{FIM_vt3} and of the \ac{SNR} scaling; moreover, it deteriorates at large angles, due to the use of the linear array.

\section{Conclusion}

This paper laid the foundation for estimating radial and transverse velocity components of a moving target using a linear ELAA, by analytically deriving the corresponding CRLBs. The results show that the accuracy of transverse velocity estimation decreases with the square of the target-antenna distance, and it is unaffected by the carrier frequency when using a half-wavelength spaced array. This behavior contrasts with the radial velocity component and also differs from near-field distance estimation. These findings suggest that transverse velocity estimation is not strictly tied to the wavefront sphericity typical of the near field, but rather depends on the geometrical arrangement of the array relative to the target's position. Hence, the array aperture plays a crucial role in achieving a satisfactory performance.

\bibliographystyle{IEEEtran}
\bibliography{IEEEabrv,Stringdefinitions,BiblioDD,biblio,WINS-Books}

\end{document}

%% file: macros.tex
\newcommand{\SNR}{\mathsf{SNR}}

\newcommand{\rect}[1] {\text{rect} \left ({#1} \right )}

\newcommand{\vr}{v_{\mathrm{r}}}
\newcommand{\vt}{v_{\mathrm{t}}}
\newcommand{\vk}{v_{k}}

\newcommand{\vrk}{v_{\mathrm{r}k}}
\newcommand{\vtk}{v_{\mathrm{t}k}}
\newcommand{\pk}{p_{k}}
\newcommand{\qk}{q_{k}}

\newcommand{\boldp} {{\bf p}}

\newcommand{\Tcp}{T_{\mathrm{cp}}}

\newcommand{\Ts}{T_{\mathrm{sym}}}

\newcommand{\Tobs}{T_{\mathrm{obs}}}

\newcommand{\Da}{D_{\mathrm{elaa}}}

\newcommand{\Ka}{K}

\newcommand{\fc} {f_{\mathrm{c}}}

\newcommand{\Ptx} {P_{\mathrm{T}}}

\newcommand{\jm} {\mathrm{j} }

\newcommand{\df} {\Delta\! f}

\newcommand{\kb}{k_{\text{B}}}

\newcommand{\Jrr}{J_{\vr\vr}}
\newcommand{\Jtt}{J_{\vt\vt}}
\newcommand{\Jrt}{J_{\vr\vt}}
\newcommand{\Jtr}{J_{\vt\vr}}

\newcommand{\ek}{\mathbf{e}_k}

%% file: acronyms.tex
\begin{acronym} 
\acro{AoA}{angle of arrival}
\acro{AWGN}{additive white Gaussian noise}
\acro{BS}{base station}
\acro{CRLB}{Cramér-Rao Lower Bound }
\acro{ELAA}{extremely large aperture array}
\acro{FIM}{Fisher information matrix}
\acro{JCS}{joint communication and sensing}
\acro{LOS}{line-of-sight}
\acro{MSE}{mean square error}
\acro{OFDM}{orthogonal frequency-division multiplexing}
\acro{SCS}{subcarrier spacing}
\acro{SIMO}{single input multiple output}
\acro{SNR}{signal-to-noise ratio}
\acro{ULA}{uniform linear array}
\end{acronym}